# How "one-size-fits-all" public works contract does it better? An assessment of infrastructure provision in Italy

Massimo Finocchiaro Castro[a, c, d], Calogero Guccio[b, c], & Ilde Rizzo [b]

[a] Department of Law, Economics and Humanities, Mediterranean University of Reggio Calabria, Italy

[b] Department of Economics and Business, University of Catania, Italy

[c] Health Econometrics and Data Group, University of York, UK

[d] Institute for Corruption Studies, Illinois State University (USA)

**Abstract**

Public infrastructure procurement is crucial as a prerequisite for public and private investments and for economic and social capital growth. However, low performance in execution severely hinders infrastructure provision and benefits delivery. One of the most sensitive phases in public infrastructure procurement is the design because of the strategic relationship that it potentially creates between procurers and contractors in the execution stage, affecting the costs and the duration of the contract. In this paper, using recent developments in non-parametric frontiers and propensity score matching, we evaluate the performance in the execution of public works in Italy. The analysis provides robust evidence of significant improvement of performance where procurers opt for design and build contracts, which lead to lower transaction costs, allowing contractors to better accommodate the project in the execution. Our findings bear considerable policy implications.

## 1. Introduction

Public procurement is a critical aspect of government operations, as it enables governments to acquire goods and services from the private sector to deliver public services (Bandiera et al., 2021).

Public infrastructure procurement is one of the most crucial domains of public procurement as it serves as a prerequisite for both public and private investments that contribute to the accumulation of economic and social capital in local territories (Papagni et al., 2021). Indeed, several studies have suggested that the development of infrastructure plays a significant role in driving economic growth (Glomm and Ravikumar, 1994), with evidence pointing towards a mutually reinforcing relationship between socio-economic and infrastructure development (Wang, 2002; Mamatzakis, 2007; Donaldson, 2018; Zolfaghari et al. 2020; Elekdag et al., 2022). Moreover, public works contracts usually involve long-term commitments, and thus, the efficient provision of infrastructure and their capability to deliver the planned benefits are severely affected by the execution problems of the underlying procurement contracts (Flyvbjerg, 2014).

Public infrastructure contract renegotiation is a critical process in ensuring that public work projects are completed efficiently, effectively, and in line with the objectives of the procurement process. Contract renegotiation can address issues such as opportunistic behavior by firms, changes in project scope, and issues of contract incompleteness, (Hart, 1995). However, despite the best efforts of both parties to agree on the terms of a contract, unforeseen events can occur during the contract period that can affect the viability of the contract (Beuve and Saussier, 2021). At the same time, public works contracts are often linked to complex transactions for which the goals can change from the original design, and it is not easy to anticipate future circumstances that may arise (Williamson, 1985). Therefore, renegotiations can be seen as necessary adjustments to fill in the blanks in the public works contract (Williamson, 1985; Grossman and Hart, 1986).

However, the performance of public works procurement mainly depends on several elements such as the design of the project, the modes of selecting contractors, the specification of the contract and the rules governing its assignment and execution.

Public works procurement has been widely investigated within the principal-agent paradigm, the crucial issue being to overcome the asymmetric information, providing incentives to the agents to meet the principal/policy-maker's objectives. In terms of regulation design, competitive tendering is fostered, the discretion of the contracting authority is reduced, and incentive contracts are provided at the implementation level, restricting the degrees of freedom for both sides of the contract, the compliance with regulation being the main objective. This one-size-fits-all approach (O'Flynn, 2007), fails to exploit the potentialities of contractual public-private relationships, disregarding the importance of trust, relational capital and relational contracts governance. Coherently, in such a theoretical perspective, a widely used system of public works delivery separates design and construction phases (i.e., DBB) with a strong tie to fixed-price contracts. Therefore, contract-winner starts construction based on a complete project created by a different subject chosen by the contracting authority. The use of DBB is advocated because the separation of builders and designer roles is claimed to provide check and balances within the projects, and, at the same time, to prevent contractors' opportunistic behaviour. However, in most projects, especially in the case of big and complex ones, the initial designs are often subject to change, leading to contractual renegotiation and resulting in large cost overruns and delays (Bajari et al., 2009).

To overcome the limitations of the above approach, which considers the contractual relation as a 'black box', it is useful to unpack such box and address its multidimensional features. The Transaction Cost Economics (TCE) offers the suitable theoretical framework, being "*one of the ''common languages'' that help to unify research across the social sciences in general and the functional areas of business administration in particular*." (Williamson, 2005, p. 37).

TCE, covering the life of the contract, minimizes the costs deriving from the fragmentation of relationships (Williamson, 1999), occurring in the transition from one contractual stage to the other (Hensher & Stanley, 2008). TCE outlines that bargaining is pervasive and does not refer only to the ex-ante contracting stage (as competitive tendering suggests) but that also ex-post stages need attention. Therefore TCE, as a governance structure, addresses the problems of ex post governance (Williamson, 2005) and it is suitable to support value creation through all the life of the contract, taking into consideration the several issues arising in contract execution. Coherently, TCE suggests the importance of fostering communication between contracting authorities and firms, enhancing the utilization of the contractor's expertise when designing the project (Hensher & Stanley, 2008) and stressing the importance of building trusting partnerships.

As Canitez & Çelebi (2018) outline, not only internal transaction characteristics but also institutional environment crucially affect the size of transaction costs, since the relationships between buyers and sellers do not occur in a vacuum: where the levels of trust and of social capital are low, transaction costs are likely to be higher, calling for vertically integrated solutions (Taylor, 2003).

Hence, TCE offers a theoretical approach to investigate whether D&B - a procurement system implying that one provider, a design-build contractor, is responsible for the design and the construction phases- is a suitable alternative solution to DBB in terms of performance. Being renegotiations the main sources of cost overruns and time delays, D&B contract should outperform the DBB system in both aspects. D&B may come up as a successful strategy not only from a quantitative point of view, as measured by costs and time, but also from a qualitative one. D&B as a form of vertical integration gets support by TCE, which emphasizes ex post governance issues in term of economizing transaction costs. Taking into consideration transaction costs through the life of the project allows to develop a synergic and collaborative public-private relationship, induced by flexible contract interpretation and enhanced information sharing, and to develop trusting relationships aimed at improving the performance of public works procurement (Cavalieri et al., 2019a; De Carolis and Palumbo, 2015).

D&B contracts, however, require adequate managerial skills and organization in the contracting authority at two levels: first, at the strategic decision-making level, to choose the most suitable design-construction proposal in relation to the work features to minimize transaction costs; second, at the implementation level, to evaluate whether the design meets the authority's specific needs and to monitor that the contractor in the execution does not take advantage of the larger degrees of freedom and flexibility related to the double roles of work's designer and builder. Being the risk shifted on the contractor, the authority is not required to spend effort in arbitrating between separated design and construction contracts and, its task might appear to be less demanding in terms of time and effort. However, the authority needs to be very careful and skilled in establishing a collaborative relationship with the contractor, based on flexibility and trust, favoring information sharing, to assess that the contractor fulfils its obligations in terms of the quality and innovative solutions contained in the bid. Such a managerial attitude and capability is crucial for a successful D&B coupled with a regulation suitable to favor private-public synergies.

Even though D&B contracting is widely used in many countries around the world, especially in the transport sector, the empirical literature on its effects on procurement performance, lacks consensus (Decarolis and Palumbo, 2015; Park and Kwak, 2017; Cavalieri et al., 2019b), and the potentialities of D&B to improve the performance of public works procurement need closer investigation. This paper, on the grounds of the TCE theoretical insights, aims at testing the above-mentioned advantages of D&B, that is whether D&B, compared with DBB, has positive effects on the performance of public works contracts, measured in terms of cost overruns and time delays.

This paper focuses on the demand side, exploring crucial aspects to grasp the above-mentioned potentialities such as the characteristics of the contracts as well as the adequacy of regulation,

On these grounds, the paper aims at addressing the following research question: what is the effect of using D&B, compared with DBB, on the performance in the execution of public works contracts, measured in terms of cost overruns and time delays?

To robustly answer this question, there are several measurement challenges that our analysis seeks to overcome. The first one concerns the choice of performance measures. While unit price comparisons can be used for standardized goods, they are not suitable for more complex contracts such as public works. Indeed, these works are heterogeneous in many dimensions, contracts are often incomplete, prices may be renegotiated, and contract execution may be delayed. For this reason, delays and cost overruns are widely used as performance indicators in the literature. Moreover, both of these indicators are proxies for Williamson's (1971) transaction or "haggling" costs, which are present whatever the reason behind the renegotiation of original contract terms and have proven to be economically relevant for complex contracts such as public works (e.g. Bajari, Houghton and Tadelis, 2014)

However, employing these indicators separately does not allow for a unified assessment of performance due to the use of D&B contracts (Decarolis and Palumbo, 2015). To overcome this limitation, we employ a unified approach of performance evaluation in public works performance first proposed by Guccio et al., (2012) and widely used in the literature in the field.

Nevertheless, this is still not enough to be able to robustly assess the relative performance between the two groups of contracts, and especially to be able to make inferences between the adoption of D&B and the performance of public works contracts. Therefore, we will empirically analyze how the adoption of D&B contract arrangements are related to the performance in public works execution by means of propensity score matching, together with non-parametric frontier (Bogetoft and Kromann, 2018). Finally, to analyze the impact of D&B use on the performance in public works execution, we estimate both Average Treatment Effect (ATE) and Average Treatment Effect on Treated (ATT) that provide robust evidence of significant improvement of performance in the execution of public works, where procurers opt for D&B contracts. The empirical investigation of our research question will be conducted using a unique dataset of public works contracts awarded in Italy between 2008 and 2014 and allows us to assess the final price and time as long as that the contract was completed by 2019.

To this end, D&B contracts in Italy appear as a promising case study. The Italian public works delivery system has a poor performance both in terms of cost overruns and, especially, large delays and is worth investigating for several reasons. The regulatory system of D&B contracts is characterized by discontinuities through time and offers room for analyzing how the private-public relationship has evolved consequently. Also, the characteristics of the Italian public works market may play a role on choosing and shaping D&B contracts. On the one hand, demand is rather fragmented and exerted by contracting authorities which are highly diversified in terms of dimension, governance– central and local governments, public enterprises, etc. – degree of specialization, quality of organization.  On the other hand, supply consists of a very high number of firms, mainly small and medium enterprises, subjected to a strict qualification system to access the market. All these features affect the management of D&B contracts as well as its impact on the performance of public procurement assessed in relation to the type of project, the characteristics of the private provider and the features of the public contracting authority.

The remainder of the study is organized as follows: the next Section illustrates the methodology and empirical strategy. Section 3 describes the Italian case and the data sample. Section 4 reports the results of the empirical analysis and, finally, Section 5 provides a discussion of results and some policy implications.

## 2. Methods and empirical strategy

In most of the empirical investigations on public works contracts performance, the ex-post assessment of the execution of public work contracts is defined in terms of either cost overruns or time delays. Its measurement is generally carried out by means of the relative excess costs and time with respect to the costs and the time agreed on in the contract with the firm (for a recent survey, see Cavalieri et al., 2019a). However, these measures have two main limitations (Guccio et al., 2012; Finocchiaro Castro et al., 2014). Firstly, they represent productivity measures, since they do not arise from a comparison with any, however determined, efficient benchmark. Secondly, considering separately the two phenomena, it does not allow for the evaluation of the overall performance of the procurer in carrying out the contract. To account for those limitations, we employ the approach first proposed by Guccio et al., (2012) to evaluate the performance in the execution of public works and largely adopted in the literature (see among several others, Finocchiaro Castro et al., 2014; Ancarani et al., 2016; Cavalieri et al., 2017; 2018; Finocchiaro Castro et al., 2018; Lisciandra et al., 2022).

Indeed, the proposed approach aims at measuring the procurers' capacity in achieving both the targeted results of time and costs, through a benchmarking of their performance, regarding as best performers those procurers that minimize the actual time and costs of execution of public works.

More specifically, to carry out benchmarking, we use a well-established and useful nonparametric methodology, Data Envelopment Analysis - DEA (Charnes et al. 1978). This is a technique generally used to estimate a production function, which is capable to handle multiple inputs and outputs without requiring a priori assumptions of a specific functional form on production technologies and the relative weighting scheme. By applying this approach, the efficiency of a generic Decision-Making Unit (DMU) like, for example, a contracting authority carrying out a public work contract, is measured by the distance between the observed input-output mix and the optimal mix located on the frontier which is the boundary of optimal production plans[1].

In what follows, we provide a short formalization of the method employed in the analysis[2]. In line with the notation used by Simar and Wilson (2008), we consider a production process using the vector of inputs $\{x = x_i, i = 1, \ldots, n\} \in \mathfrak{R}_+^N$ that is used to produce a vector of outputs $\{y = y_s, s = 1, \ldots, m\} \in \mathfrak{R}_+^M$. The production process is constrained by the production possibility set $\Psi$, which is the set of physically attainable points $(x, y)$ given by:

$$\Psi = \{(x,y) \in \mathfrak{R}_+^{N+M} | = (x,y) \text{ is feasible } \} \qquad (1)$$

The efficiency of a generic DMU is measured by the distance between the observed input-output mix and the optimal mix located on the frontier of $\Psi$, which is the boundary of optimal production plans.

The single DMU efficiency score, as defined by Debreu (1951) and Farrell (1957) in the input-oriented case, is:

$$\lambda(x,y) = \inf\{\lambda | (\lambda x, y) \in \Psi\} \qquad (2)$$

where a value of $\lambda(x,y) < 1$ measures the radial distance of the DMU from the full efficient frontier and a value of $\lambda(x,y) = 1$ means that the DMU is fully efficient. Being $\Psi$ the frontier and $\lambda(x,y)$ unknown, they should be estimated from a sample of i.i.d. observations $\mathcal{X}_n = \{(x_i, y_i), i = 1, \ldots, n\}$.

---

[1] For technical details on non-parametric estimators, see, among the others, Simar and Wilson (2008).
[2] For more details, see Simar and Wilson (2008).

The DEA estimator assumes the convexity of the hull and, thus, under the hypothesis of constant returns to scale (CRS), can be defined as:

$$\widehat{\Psi}_{DEA} = \{(x,y) \in \mathbb{R}_+^{N+M} | y \leq \sum_{i=1}^{n} \gamma_i x_i; \; x \geq$$
$$\sum_{i=1}^{n} \gamma_i y_i, \text{for } (\gamma_1, \dots, \gamma_n) \text{ such that } \gamma_i \geq 0, i = 1, \dots, n \} \tag{3}$$

A DEA non-parametric estimator of the efficiency scores can be calculated by replacing the true production set $\Psi$ in (2) with the estimator $\widehat{\Psi}_{DEA}$:

$$\hat{\lambda}_{DEA}(x,y) = \inf\{\theta | (\theta x, y) \in \widehat{\Psi}_{DEA}\} \tag{4}$$

where, by construction, $\hat{\lambda}_{DEA}(x,y) \leq \lambda(x,y)$ (Simar and Wilson, 2008).

However, to be able to robustly assess the relative performance between the two groups of contracts, and especially to be able to make inferences between D&B adoption and performance of public works contracts, we need to control for other environmental factors that potentially affect the performance (Simar and Wilson, 2008).

We use different empirical strategies to assess the robustness of our results. Specifically, we first use a fully non-parametric approach to assess relative performance considering the role of environmental factors. First, we test for the presence of a common or separate frontier (Simar and Wilson, 2007) between the D&B and DBB contracts using the results of Daraio et al. (2018) and Simar and Wilson (2020) and, to be conservative, we employ the efficiency estimates of both common and separate frontiers in the second stage analysis. Then, to test whether the average performance between DDB and D&B is significantly different, we use both kernel density functions and the equivalence test proposed by Kneip et al., (2016). However, this approach, although robust, has the limitation of not allowing us to evaluate for unobservable factors that could potentially affect performance.

Thus, we employ a matching estimator proposed by Bogetoft and Kromann (2018) that allows us to robustly assess the role of D&B in generating better performance for contracting authorities. Indeed, matching creates a "quasi-randomized" experiment by matching the two groups of public works contracts based on observable information (Caliendo and Kopeinig, 2008). Bogetoft and Kromann (2018) prove that using matching together with DEA ensures sub-sample homogeneity and eliminates sample size bias. The underlying idea is to create a non-treatment counterfactual group that matches the treated contracts in a series of covariates. Ideally, all the relevant differences in the performance of the control and treatment groups are captured by their observed characteristics. As such, conditioning on the observable's covariates, both observed and unobserved differences between the treatment and control groups are eliminated, thereby essentially simulating random assignments. However, if the number of covariates is large, matching each characteristic of the contract is difficult and the chances of finding an exact match are reduced. To overcome this curse of dimensionality, the authors suggest employing the Propensity Score Matching (PSM) first proposed by Rosenbaum and Rubin, (1983).

In our estimates, the PSM is calculated for each contract in the two groups of D&B and DBB, and the best match is found. Of course, this approach is not as good as pure randomization, as we can only match on observed characteristics. However, if most of the characteristics related to the treatment assignment are observed, one can be confident that comparisons of the outcomes across the two groups will reflect treatment effects. Specifically, PSM matches every D&B with the closest DBB contracts and *vice versa*. The crucial step in the matching procedure is the choice of covariates. In our empirical exercise, we will employ those most used in the related literature (Decarolis and Palumbo, 2015)[3].

Regarding the choice of the PSM method, following Bogetoft and Kromann (2018) we employ the nearest neighbor (NN) matching. In the application of the NN estimator, D&B contracts (treated) are matched with DDB (as the control group) with the NN estimated propensity scores. However, to ensure the robustness of our findings we employ as an alternative method the genetic multivariate matching search algorithm proposed by Diamond and Sekhon (2013).

Since the propensity score is the probability of receiving treatment (in our case, D&B), researchers can choose any discrete choice model. In our case, we will employ a logit model. After estimating the propensity score, before applying a chosen matching estimator, it is necessary to conduct a balancing test[4]. For this purpose, we applied the procedure of Becker and Ichino (2002).

---

[3] See the next section for a description of the covariates used.
[4] The purpose of a balancing test prior to matching (stratification test) is to test how well the estimated propensity score was able to balance the covariates.

After the PSM estimate, to analyze the impact of D&B use on performance in public works execution we estimate both average treatment effect (ATE) and average treatment effect on treated (ATT). In simple terms, the ATE enables us to estimate the difference in expected average outcomes after D&B and DBB treatments, while the ATT enables us to estimate the difference between expected outcome values with and without D&B for those public works that use the D&B contract. In this way, we believe that we can robustly assess if the use of D&B contract can generate higher performance for procurers. In fact, ATE allows us to compare the outcome in each contract execution (i.e., the estimated efficiency score) with the average outcome of the matched contracts and estimate the average effect of D&B as the average of these comparisons[5]. The ATT allows us to assess the difference in outcomes for the treated public works contract (i.e., D&B) with and without treatment.

### 3. The Italian case and data sample

#### 3.1 Public infrastructure procurement in Italy

As mentioned previously, our empirical assessment takes the Italian regulatory framework as a case study and focuses on the role of contracting authorities being, at the same time, the subjects most interested in D&B contracts and the public entities with the greatest risk of capture by providers (Gori et al., 2022). There is a wide academic and political debate on the poor performance of public works contracts, with cost overruns and time delays being identified as the main causes almost everywhere (OECD, 2013) and in Italy, too (Cavalieri et al., 2019b).

In Italy, as a tool to mitigate the low performance of public works contracts, the policymaker has extensively relied upon heavy regulation. No attempt is made here to analyze in depth the Italian legislation and only few key features will be recalled. It is worth noting that since the beginning of the nineties to address asymmetric information problems, both in terms of adverse selection and moral hazard, the law assigns great emphasis to the design phase as a tool to reduce the risk of renegotiation, aims at promoting competition to select the most convenient provider and to reduce bureaucratic discretion, uses fixed price contracts to prevent the opportunistic behavior of private contractors and provides a detailed regulation for qualifying firms to access the public works market. Moreover, within the legislation aimed at preventing corruption and enhancing transparency in the public sector, specific provisions are offered to improve procurement integrity and transparency, under the supervision of the National Anticorruption Authority (ANAC). Many changes have been introduced through time, overall generating high instability of the Italian procurement sector and great uncertainty for both public and private actors.

---

[5] See Caliendo and Kopeinig, (2008) for technical details.

Among the various aspects of Italian regulation, here attention is focused on the former stage of the procurement process, namely on the design of the project. We do not investigate the choice regarding the internal or external assignment of the project design (Cavalieri et al., 2019b) but we concentrate on a specific aspect of the latter, that is whether the final stage of design is separated from the construction phase (DBB) or whether the final design and the construction phases are in the responsibility of the same provider (D&B).

DBB was adopted in the nineties reform as a rule to empower the contracting authority and prevent opportunistic behaviors deriving from the assignment of the roles of the work's designer and builder to the same provider. Since then, it has been highly debated and subjected to many changes.

In theory, such a debate can be interpreted as the search for a compromise between competition and fixed price contracts, on the one hand, leading to strengthening DBB, and the need to speed up the completion of public works contracts and reduce time delays, on the other hand, favoring D&B solutions.

In practice, D&B has experienced many up and downs. In 2002 limited exceptions to DBB were introduced and in 2006, in line with EU legislation, the possibility of using D&B contracts was further expanded. Finally, in 2016 the separation between the design phase and the execution phase, has been reintroduced and stressed. This restrictive approach, however, has not lasted long. In fact, as a response to the increasing and generalized concern for the length of public works contracts implementation and for the related delays, since 2019, Government has introduced many temporary rules to suspend the effects of the 2016 law preventing the use D&B contracts, firstly, until the end of 2020 and, subsequently, until 2021 and finally until June 2023.

Recently, the tendency to allow faster and simple procedures has been reinforced to face on the one hand, the economic crisis generated by the Covid-19 pandemic and, on the other hand, the expected increase in resources dedicated to infrastructural investments deriving from the National Recovery and Resilience Plan (NRRP) requiring strict time commitments (Baltrunaite et al., 2021). In 2021, an 'extended' D&B has been introduced, only for public works contracts related to PNRR, or anyway financed out of EU funds. Thus, the suspension of the rule preventing D&B contracts has been extended until the end of 2023. While there is some evidence that D&B has positive effects on the time completion of public works contracts, and it has been estimated that the length is reduced by 8%, the effects on the quality of works are an open question (Gori et al., 2022).

The Italian regulatory system of D&B contracts being characterized by the above-mentioned discontinuities through time offers room for analyzing how the private-public relationship can be managed in a strategic way to increase the performance in public works contracts, to meet contingent socio-political objectives.

### 3.2 Data

Our main dataset is provided by the Observatory of the Public Contracts. at the ANAC. Our sample virtually cover all contracts with reserve price between 40,000 and 200,000 euro awarded between 2008 and 2014 and completed at the last date of access to data. In fact, all contracts in our sample have been concluded, leading to the execution of public work (e.g. Decarolis and Palumbo, 2015; Cavalieri et al., 2019a). Our choice to examine contracts awarded between 2008 and 2014 allows us to assess the final price and the time of completion of the contract and minimizes the risk of censoring bias[6]. Furthermore, we also limit the analysis to contracts with a reserve price below 200,000 euros for two reasons. Firstly, it is due to the fact that throughout the period covered by our data, Italian regulations allowed contracting authorities to freely choose the use of a D&B only below this threshold[7]. The second reason is that this choice allows us to compare our findings with the results obtained by Decarolis and Palumbo (2015) not employing the DEA.

---

[6] Censoring bias results from the presence of truncated data. In fact, the more recent the contracts, the higher the proportion of incomplete projects is (Decarolis and Palumbo, 2015). Considering completed contracts only, we focus on works that are as likely to be completed to avoid censoring bias.

In our dataset, for each contract, we have detailed information about the type of contracting authority, the procedure and the selection criterion used to award the contract, the number of bidders, and the identity of the winning bidder. The data also includes information on public work outcomes, such as the initial project value (i.e. reserve price)[8], the winning rebate and the total effective costs, the expected and effective contractual time[9].

## 4   Empirical findings

### 4.1 Preliminary findings

Given that our empirical approach involves several steps (i.e. at the first stage we identify comparable public works groups through the PSM, at the second stage we estimate the efficiency frontiers in contract execution, and finally estimate ATE and ATT values), for convenience, we present first the data used in the PSM.

Being the contracts in the two groups of D&B and DBB not randomly assigned, the groups may differ considerably in terms of "pre-treatment" characteristics. This can seriously hamper the validity of conclusions on treatment effects found by comparing D&B and DBB contracts. To create a useful matched control group, the variables included in the logistic regression should ideally be all the variables that potentially confound the treatment effect. Following Bogetoft and Kromann (2018), we can consider in general terms potential confounders: covariates available prior to treatment assignment that may influence the treatment decision and may influence the outcome (in our case the performance in public works execution). Furthermore, all covariates relating to factors occurring after the treatment choice (in our case D&B or DBB) must be excluded.

---

[7] Legislative Decree n. 163/2006 that transposed the European Directive n. 2004/18/EC.
[8] The terms reserve price or base price are indifferently employed to indicate the value of the project.
[9] Employing DEA estimator and PSM estimates, we have checked for data completeness and outliers.

However, if the number of covariates is large, matching each characteristic of the contract is difficult and the chances of finding an exact match are reduced. Therefore, we choose the minimum number of covariates to optimize our matching problem. To this aim, in line with the previous literature, the set of covariates for PSM includes controls for the reserve price, type of public work (new build or restoring), type of contracting authorities (i.e. municipality, province, region, etc.) and award procedure (negotiated or auction precedures)[10].

In the second step, to assess the performance we use an input-oriented approach assuming that for a given target of time and cost, agreed on in the contracts, the most efficient procurers are the ones that minimize the actual time and costs (Guccio et al., 2012). Furthermore, each public work contract is treated as a separate DMU with its own input and output values[11].

As our purpose is to assess *ex ante* whether there are latent or unobservable factors in the comprehensive dataset, that make the two groups non-comparable, following Daraio et al. (2018) and Simar and Wilson (2020), we test the separability of the frontier with respect to discrete environmental variables in the comprehensive dataset. Specifically, we assume that the choice between D&B or DBB can be considered a discrete environmental variable that potentially could affect the performance of the two groups of contracts[12].

We obtain a statistic $\tau = 3.4307$ and a KS = 0.5148 with corresponding *p-values* less than 0.0001 in both cases, indicating that in the comprehensive dataset there are latent factors that impact the performance of the two groups of contracts making the comparative evaluation potentially biased. [13] This first result confirm that it is necessary to employ a PMS, before estimating the efficiency with DEA, to properly evaluate the performace. Therefore in the next Section we first perform the PSM estimates, which ensures that we can identify comparable groups of contracts, and then we assess the relative performance in contract execution[14].

---

[10] We also tested other potential covariates, such as public works category or region of public work location, but they were either not significant or failed to pass the property balance rule.

[11] Table A.2 in the Appendix A reports descriptive statistics for both the full sample and the two subsamples of contracts using D&B and DBB contracts respectively.

[12] Descriptive statistics of the estimates obtained in the comprehensive dataset and under several frontier assumptions are provided in Tables A.3 and A.4 in Appendix A.

[13] Moreover, the low comparability between the two groups of contracts is also quite evident from a simple inspection of the estimates in Tables A.3 and A.4 in Appendix A.

### 4.2 Assessing the performance

The propensity score was estimated by a logit model and the results are reported in Table 1.

*<< Table 1 about here>>*

The results indicate a statistically significant positive effect of reserve price and new building in the use of D&B contracts. In addition, D&B is more frequently employed with auctions and compared with the reference group given by municipalities. D&B contracts are less frequently employed by all contracting authorities except central government. However, in the matching approach, the focus is not on the estimated coefficients but rather on whether covariates between matched pairs of treated and untreated contracts are balanced, given the estimated propensity scores. Furthermore, the literature on matching suggests the inclusion of also the covariates that are statistically insignificant because their inclusion does not increase bias in subsequent matching estimations (Caliendo and Kopeinig, 2008).

As PSM we employ both NN and genetic multivariate matching search algorithm proposed by Diamond and Sekhon (2013). Figure 1 provides clear evidence of the distributions of propensity scores by group of contracts employing the NN and genetic PSM, respectively. Firstly, in both algorithms there are no treated units that are not matched; secondly, the group of treated (i.e., D&B) and of matched untreated (i.e., DBB) appear quite similar in terms of distribution, whereas the contracts discarded from matching show a different distribution of propensity scores to that of treated.

---

[14] For nonparametric estimation we employ the FEAR package developed in R by Wilson (2008). PSM estimates were performed in R with the packages MatchIt (Stuart et al., 2011) and Matching developed by Sekhon (2011).

The latter result is clearly shown in Figure 2, which reports the standardized mean differences on the covariates of the matched contracts. More specifically, Figure 2 shows the balance on the covariates after matching (in terms of standardized mean difference) according to the matching method; the benchmark (the unmatched contract) is in each case outperformed by any matching method, but the genetic method is the most effective. Thus, to ensure the robustness of our findings, we will use the contracts identified with both matching methods to estimate the relative performance in public works execution and then calculate the ATE and ATT.

Table 2 shows the descriptive statistics of the efficiency scores computed on the two samples of matching contracts obtained with the NN and the genetic multivariate matching, respectively[15]. As might be expected, the average efficiency level is rather modest and lies within the range of the efficiency estimates obtained on the whole sample[16]. The results reported in Table 2 show that, being the average efficiency score between 38% to 42%, each contracting authority can reduce, on average, both actual time and costs proportionally by 58% to 62% given the target value (that is, the time and costs agreed on in the contract)[17].

However, the picture change if we look at D&B and DBB contracts. Contracts employing D&B show an efficiency level between 43% to 44%, whereas DBB contracts reach efficiency levels between 38% to 41%[18]. The first descriptive analysis seems to confirm what we obtained looking at the full sample in Appendix A; *i.e.* that D&B contracts appear to perform comparativelly better than DBB ones in the execution of public works[19]. Furthermore, these findings are in line with the results of Bogetoft and Kromann (2018) that, in the absence of matching, a large part of the difference in terms of efficiency between the groups is the result of selection bias.

---

[15] The descriptive statistics of the input and output variables in the two samples are shown in Table A.5 in Appendix A.

[16] See Tables A.2 and A.3 in Appendix A.

[17] It is important to highlight that the fully efficient contracts (*i.e.,* those on frontiers) are not necessarily the ones that fulfil simultaneously time and cost efficiency. In fact, in the input-oriented CRS model (Charnes et al.,1978) employed here, the efficiency score measures the radial contraction in the actual achievements of cost and time objectives needed to attain the contract target in relative terms. Thus, it identifies the best performing contracts, in the relevant trait of the bi-dimensional frontier, as the ones that minimize the "distance" of actual achievements from the targets. This implies that the best performing contracts could still exhibit a relatively inefficient performance in one of the targets (i.e., time and costs) agreed on in the contract.

<< *Table 2 about here*>>

However, this preliminary result is not sound enough to assess robustly the impact of the use of D&B on performance in the execution of public works. For this purpose, in Table 3 we report the estimates of both the average treatment effect (ATE) and the average treatment effect on treatment (ATT) on the NN and genetic matching. These data are consistent between the two matching estimators. Table 3 shows a positive and significant impact of D&B on performance in the execution of public works for both ATE and ATT. More specifically, the results in Table 3 show that the average effect of using the D&B contract is an increase in performance of 5.7 per cent in the NN matching and 3.6 per cent for genetic multivariate matching. Looking at the ATT, the results in Table 3 show that for both matching estimators the effect of using D&B contracts is positive and statistically different to zero, but not much different than the ATE. Summing up our empirical findings provide robust evidence of significant improvement of performance in the execution of public works, where procurers opt for D&B contracts.

<< *Table 3 about here*>>

## 5   Discussion, limitation and conclusion

---

[18] The kernel density estimates reported in Figure A.1 in the Appendix A further confirm the indications emerging from Table 3.

[19] The differences between the two groups are even larger for the full sample both for common and for separated frontier estimates. See Tables A.2 and A.3 in the Appendix A.

The key findings of this study highlight that, all things being equal, the use of D&B contracts may lead to higher performance in the execution of public works.

The results of a logistic regression indicate that the contracting authorities were most likely to use D&B contract on public works related to new construction and with higher reserve prices.

The main strength of this study is the capability to assess efficient performance in public works execution by integrating DEA with PSM; this leads to a more robust efficiency assessment on comparable contracts. The PSM procedure allows for the creation of paired groups of contracts and enables the integration of efficiency scores in this pairing process. The integration provides better identification in public works efficiency by creating more comparable groups. In addition, efficiency sensitivity analysis using a genetic algorithm underscores the validity and robustness of the results.

Using both the NN and the multivariate matching genetic algorithm proposed by Diamond and Sekhon (2013) the results of the ATE and ATT regressions provide clear and robust evidence of significant improvement of performance in the execution of public works, where procurers opt for D&B contracts.

This result has relevant policy implications both on the demand and supply side. Firstly, to grasp the positive effects of D&B on public works performance, the procurer needs to follow a managerial strategy aimed at enhancing its capability to clearly identify its needs and final objectives, to share information with the potential contractors, to monitor the implementation of the contract through a well-defined set of operational indicators that are simple, easily manageable but capable of promptly signaling any elements of risk – red flags - that require immediate correction. In other words, the aspects related to the organizational structure are crucial to enhance the procurer's performance and prevent the 'capture' by the contractor, the policy implication being a regulation system stressing the qualification of the buyer and its ability in managing and monitoring synergic public-private relationships. This regulatory feature deserves special attention when both central and local governments are involved, leading to different procurement department organizations (Patrucco et al., 2017) considering the difficulties of small local contracting authorities.

Conversely, on the supply side, it is important to provide the right incentives to private contractors to induce them to build a positive relationship with the public buyer, without exploiting their information advantage. To this end, focusing on reputation is widely recognized in the literature as an important and effective tool (De Carolis et al, 2016), a policy implication being to enhance the role of reputation in regulating the qualification of firms to enter the public works market.

Providing clear and robust evidence that D&B contracts significantly improve the performance in the execution of public works, this paper also suggests that in presence of strict time commitments for the completion of public works, as those deriving by the NRRP, the most recent Italian regulation favoring the use of D&B, finds empirical support. However, to be effective such a choice needs to be accompanied by appropriate reforms aimed at improving the managerial skills and the organization of contracting authorities, designed on their characteristics, and the quality of suppliers, according to the lines indicated above.

Also, we acknowledge that this study has also some limitations. First, it is reasonable to assume that there are some factors affecting the decision to use D&B contracts other than those examined here. Thus, failure to consider these factors could potentially lead to a bias in our matching and not allow for a robust relationship.

Second, although the study used a robust approach when looking at the comparative efficiency, the sample of D&B contracts was much smaller than the sample of DDB ones. While the availability of a large sample of DBB contracts allowed us to robustly identify matching with D&B contracts, this difference between the two groups could still have uncontrolled effects in the execution phase. However, the efficiency estimates on full sample reported in Appendix A underscore the validity and robustness of the findings. Furthermore, we note that our estimates on ATE, and ATT are rather conservative compared to what we would have obtained by using performance estimates on the whole sample of public works[20].

Thirdly, our sample of works is limited to small-scale infrastructures. The reasons for this have been explained above and are mainly empirical in nature.

---

[20] The results of these additional estimates are available from the authors upon request.

However, we admit that this choice limits the possibility of generalizing the results reported here to larger infrastructures for which future research is needed.

Finally, although our empirical findings are robust with respect to several checks, the conclusions made in this study are still tentative, and several issues remain open to scrutiny and further research is needed for a thorough understanding of the phenomenon of contract renegotiation that severely affects the efficient provision of public infrastructure.

# TABLES AND FIGURES

**Figure 1** - Distribution of propensity scores matching.

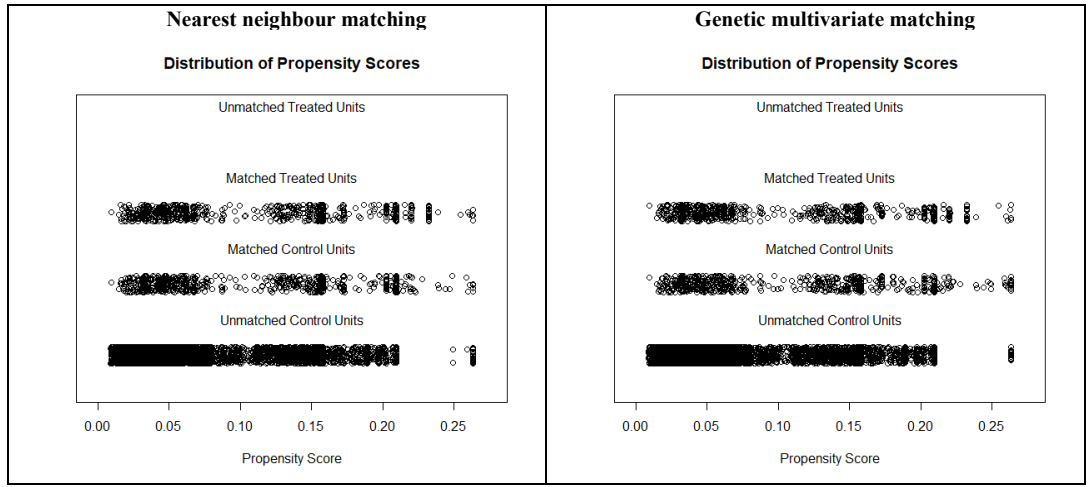

*Source:* our elaboration on data provided by the Observatory of the Public Contracts - ANAC.

*Note:* The figure reports the distributions of propensity scores matching of NN (on the left) and genetic multivariate matching search algorithm (on the right) proposed by Diamond and Sekhon (2013). Public works contracts with base prices between €40,000 and €200,000, awarded by Italian contracting authorities in the period between 2008 and 2014, and completed at the last date of data availability.

**Figure 2** - Standardized mean differences

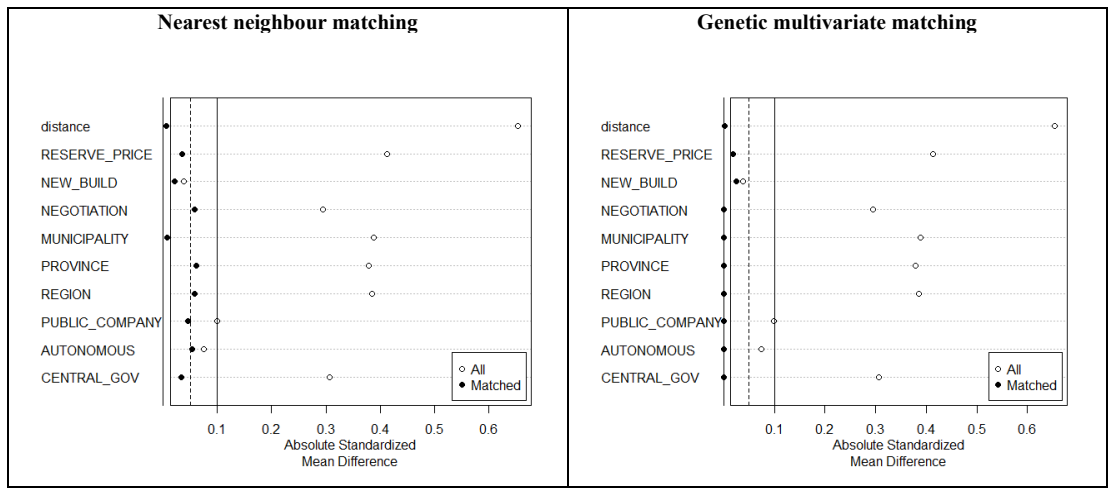

*Source:* our elaboration on data provided by the Observatory of the Public Contracts - ANAC.

*Note:* The figure reports the absolute standardized mean difference on the covariates of the matched contracts using NN (on the left) and genetic multivariate matching search algorithm (on the right) proposed by Diamond and Sekhon (2013). Public works contracts with base prices between €40,000 and €200,000, awarded by Italian contracting authorities in the period between 2008 and 2014, and completed at the last date of data availability.

**Table 1** – DEA (CRS) efficiency estimates

| Sample | Obs | Mean | SD | Min | Max |
|---|---|---|---|---|---|
| **Nearest neighbour matching** | | | | | |
| Matched contracts | 1,250 | 0.4049 | 0.1950 | 0.0802 | 1.0000 |
| DBB contracts | 625 | 0.3805 | 0.1925 | 0.0802 | 1.0000 |
| D&B contracts | 625 | 0.4292 | 0.1946 | 0.1268 | 1.0000 |
| **Genetic multivariate matching** | | | | | |
| Matched contracts | 1,250 | 0.4229 | 0.1700 | 0.0841 | 1.0000 |
| DBB contracts | 625 | 0.4063 | 0.1646 | 0.0841 | 1.0000 |
| D&B contracts | 625 | 0.4395 | 0.1739 | 0.1297 | 1.0000 |

*Source:* our elaboration on data provided by the Observatory of the Public Contracts - ANAC.

*Note*: The table reports descriptive statistics of DEA (CRS) efficiency estimates for the matched contracts using NN and genetic multivariate matching search algorithm proposed by Diamond and Sekhon (2013). Public works contracts with base prices between €40,000 and €200,000, awarded by Italian contracting authorities in the period between 2008 and 2014, and completed at the last date of data availability.

**Table 2** - Average treatment effects (ATE) and Average treatment effects on the treated (ATT)

| Outcome - DEA (CRS) scores | (1) Nearest neighbour matching | (2) Genetic multivariate matching |
|---|---|---|
| **Average treatment effects (ATE)** | | |
| D&B contracts | 0.057*** | 0.036*** |
| | (0.010) | (0.009) |
| n. treated | 625 | 625 |
| n. controls | 625 | 625 |
| Obs. | 1,250 | 1,250 |
| **Average treatment effects on the treated (ATT)** | | |
| D&B contracts | 0.059*** | 0.035*** |
| | (0.012) | (0.011) |
| n. treated | 625 | 625 |
| n. controls | 625 | 625 |
| Obs. | 1,250 | 1,250 |

*Source:* our elaboration on data provided by the Observatory of the Public Contracts - ANAC.

*Note*: Table reports the Average treatment effect (ATE) and the Average treatment effect on the treated (ATT) using NN and genetic multivariate matching search algorithm proposed by Diamond and Sekhon (2013). Public works contracts with base prices between €40,000 and €200,000, awarded by Italian contracting authorities in the period between 2008 and 2014, and completed at the last date of data availability. Bootstrapped robust standard errors in parentheses. Significance: $* p < 0.1$, $** p < 0.05$, $*** p < 0.01$.

# APPENDIX A

This appendix reports additional statistics and empirical estimates to those used in the paper.

**Table A.1 -** Descriptive statistics for the matching confounders

| Variable | Meaning | Obs. | Mean | SD | Min | Max |
|----------|---------|------|------|-----|-----|-----|
| RESERVE PRICE | Contract reserve price in thousand euro | 9,394 | 157.50 | 43.19 | 40.00 | 200.00 |
| NEW_BUILD | Dummy for new building | 9,394 | 0.39 | 0.41 | 0.00 | 1.00 |
| NEGOTIATION | Dummy for negotiation procedures | 9,394 | 0.45 | 0.50 | 0.00 | 1.00 |
| MUNICIPALITY | Dummy=1 if procurer municipality | 9,394 | 0.50 | 0.50 | 0.00 | 1.00 |
| PROVINCE | Dummy=1 if procurer province | 9,394 | 0.12 | 0.33 | 0.00 | 1.00 |
| REGION | Dummy=1 if procurer region | 9,394 | 0.10 | 0.30 | 0.00 | 1.00 |
| PUBLIC COMPANY | Dummy=1 if procurer public company | 9,394 | 0.11 | 0.32 | 0.00 | 1.00 |
| AUTONOMOUS | Dummy=1 if procurer autonomous entity | 9,394 | 0.03 | 0.18 | 0.00 | 1.00 |
| CENTRAL | Dummy=1 if procurer central government | 9,394 | 0.13 | 0.34 | 0.00 | 1.00 |

*Source:* our elaboration on data provided by the Observatory of the Public Contracts - ANAC.

*Note*: The table reports descriptive statistics of public works contracts with base prices between €40,000 and €200,000, awarded by Italian contracting authorities in the period between 2008 and 2014, and completed at the last date of data availability. Monetary values in thousand euros at current prices.

**Table A.2 -** Results of logit estimates in PSM

| Variables | (1) |
|-----------|-----|
| CONSTANT | -3.0040*** |
| | (0.2341) |
| RESERVE_PRICE | 0.0000*** |
| | (0.0000) |
| NEW_BUILD | 0.3004*** |
| | (0.0967) |
| NEGOTIATION | -0.3543*** |
| | (0.0925) |
| PROVINCE | -1.0777*** |
| | (0.1146) |
| REGION | -1.6229*** |
| | (0.2073) |
| PUBLIC_COMPANY | -0.9501*** |
| | (0.1648) |
| AUTONOMOUS | -1.0449*** |
| | (0.2876) |
| CENTRAL_GOV | 0.3446*** |
| | (0.1228) |
| No of obs. | 9,394 |
| Log-likelihood | 334.22 |
| Pseudo $R^2$ | 0.0727 |

*Source:* our elaboration on data provided by the Observatory of the Public Contracts - ANAC.

*Note*: Table reports the logit estimates for the use of D&B contracts for PSM. Public works contracts with base prices between €40,000 and €200,000, awarded by Italian contracting authorities in the period between 2008 and 2014, and completed at the last date of data availability. Cluster robust standard reported in parentheses. Significance: $*$ p < 0.1, $**$ p < 0.05, $***$ p < 0.01. The results indicate a statistically significant positive effect of reserve price and new building in the use of D&B contracts. In addition, D&B is more frequently employed with auctions and compared with the reference group given by municipalities, D&B contracts are less frequently employed by all contracting authorities except central government.

**Table A.3** – Descriptive statistics of DEA inputs and outputs for the full sample

| Variables | Definition | Full sample | | | DDB contracts | | | D&B contracts | | |
|---|---|---|---|---|---|---|---|---|---|---|
| | | Obs. | Mean | St. Dev. | Obs. | Mean | St. Dev. | Obs. | Mean | St. Dev. |
| | | | | **Inputs** | | | | | | |
| A_COST | Actual cost of public work completion | 9,394 | 149.868 | 71.951 | 8,769 | 149.190 | 73.360 | 639 | 159.378 | 47.018 |
| A_TIME | Actual time of public work completion | 9,394 | 153.520 | 139.829 | 8,769 | 152.399 | 139.356 | 639 | 169.243 | 145.513 |
| | | | | **Outputs** | | | | | | |
| W_BID | Agreed cost of public work completion | 9,394 | 131.002 | 42.116 | 8,769 | 130.124 | 42.144 | 639 | 143.314 | 39.773 |
| P_TIME | Planned time of public work completion | 9,394 | 103.046 | 89.777 | 8,769 | 101.544 | 88.599 | 639 | 124.128 | 102.703 |

*Source:* our elaboration on data provided by the Observatory of the Public Contracts - ANAC.

*Note:* The table reports the time and cost value of public works contracts with base prices between €40,000 and €200,000, awarded by Italian contracting authorities in the period between 2008 and 2014, and completed at the last date of data availability. Monetary values in thousand euros at current prices. Time in days.

**Table A.4** – Efficiency estimates, common frontiers between groups for the full sample

| Variable | Obs | Mean | SD | Min | Max |
|---|---|---|---|---|---|
| | | **Full sample** | | | |
| DEA_SCORE | 9,394 | 0.3282 | 0.1644 | 0.0333 | 1.0000 |
| | | **DBB contracts** | | | |
| DEA_SCORE | 8,769 | 0.3195 | 0.1616 | 0.0333 | 1.0000 |
| | | **D&B contracts** | | | |
| DEA_SCORE | 625 | 0.4502 | 0.1558 | 0.1340 | 1.0000 |

*Source:* our elaboration on data provided by the Observatory of the Public Contracts - ANAC.

*Note:* The table reports descriptive statistics of DEA CRS efficiency estimates assuming common efficiency frontier for the full sample of public works contracts with base prices between €40,000 and €200,000, awarded by Italian contracting authorities in the period between 2008 and 2014, and completed at the last date of data availability.

**Table A.5** – Efficiency estimates, separated frontiers between groups for the full sample

| Variable | Obs | Mean | SD | Min | Max |
|---|---|---|---|---|---|
| | | **Full sample** | | | |
| DEA_SCORE | 9,394 | 0.3380 | 0.1780 | 0.0307 | 1.0000 |
| | | **DBB contracts** | | | |
| DEA_SCORE | 8,769 | 0.3194 | 0.1618 | 0.0307 | 1.0000 |
| | | **D&B contracts** | | | |
| DEA_SCORE | 625 | 0.6000 | 0.1888 | 0.1452 | 1.0000 |

*Source:* our elaboration on data provided by the Observatory of the Public Contracts - ANAC.

*Note:* The table reports descriptive statistics of DEA CRS efficiency estimates assuming separated frontier for the full sample of public works contracts with base prices between €40,000 and €200,000, awarded by Italian contracting authorities in the period between 2008 and 2014, and completed at the last date of data availability.

**Table A.6** – Descriptive statistics of DEA inputs and outputs for the matched samples

| Variables | Definition | Nearest neighbour matching | | | Genetic multivariate matching | | |
|---|---|---|---|---|---|---|---|
| | | Obs. | Mean | St. Dev. | Obs. | Mean | St. Dev. |
| A_COST | Actual cost of public work completion | 1,250 | 157.411 | 53.115 | 1,250 | 162.442 | 65.378 |
| A_TIME | Actual time of public work completion | 1,250 | 156.900 | 137.443 | 1,250 | 156.825 | 136.307 |
| W_BID | Agreed cost of public work completion | 1,250 | 142.864 | 39.590 | 1,250 | 144.889 | 39.092 |
| P_TIME | Planned time of public work completion | 1,250 | 116.337 | 116.337 | 1,250 | 113.368 | 94.150 |

*Source:* our elaboration on data provided by the Observatory of the Public Contracts - ANAC.

*Note:* The table reports the time and cost value for the sample of the matched contracts using NN and genetic multivariate matching search algorithm (on the right) proposed by Diamond and Sekhon (2013). Monetary values in thousand euros at current prices. Time in days.

**Figure A.1** - Kernel density estimates

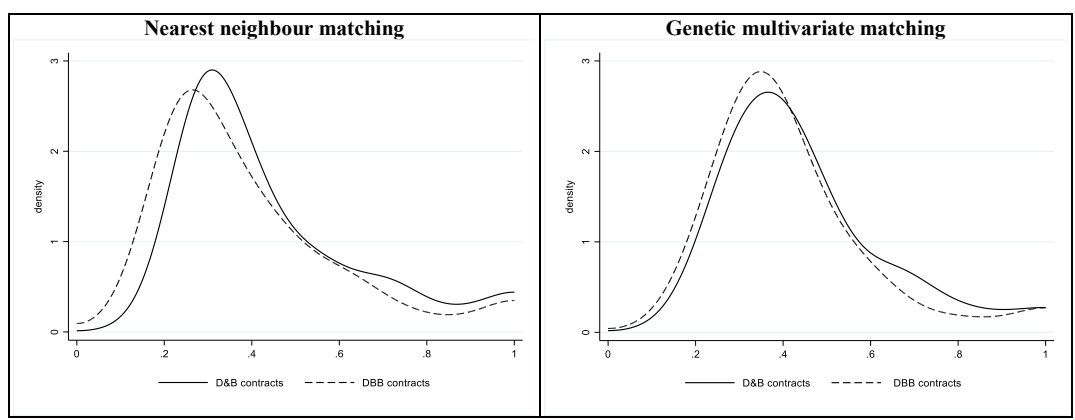

*Source:* our elaboration on data provided by the Observatory of the Public Contracts - ANAC.

*Note:* The figure reports the kernel density estimates of DEA efficiency scores under CRS assumption on the matched contracts using NN (on the left) and genetic multivariate matching search algorithm (on the right) proposed by Diamond and Sekhon (2013). The kernel density functions are derived from DEA efficiency scores using a univariate kernel smoothing distribution and the appropriate bandwidth (Simar and Wilson, 2008). Public works contracts with base prices between €40,000 and €200,000, awarded by Italian contracting authorities in the period between 2008 and 2014, and completed at the last date of data availability.